# Curvature effect induces topological phase transitions in two - dimensional topological superconductor

Huan-Wen Lai[a,b] *, Meng-Chien Wang[b], Ching-Ray Chang[a,b] and Seng-Ghee Tan[c]

[a] *Quantum information center, Chung Yuan Christian University, Taoyuan, Taiwan.*
[b] *Department of Physics, National Taiwan University, Taipei, Taiwan.*
[c] *Department of Optoelectric Physics, Chinese Culture University, Taipei, Taiwan.*

***Abstract***

Recently, topological superconductors have emerged as a pivotal frontier in condensed matter physics, owing to the exotic properties of Majorana quasiparticles residing on edges, surfaces, and vortex cores. In this work, we investigate a two-dimensional s+p wave noncentrosymmetric superconductor (NCS) with Rashba spin-orbit coupling (SOC) on a cylindrical manifold. By employing a thin-layer quantization approach, we derive the effective Bogoliubov-de Gennes (BdG) Hamiltonian to elucidate the interplay between surface curvature and topological phase transitions. Our analytical and numerical results demonstrate that the curvature-induced geometric potential (da Costa potential) and the spatial reconfiguration of spin-momentum locking effectively modulate the superconducting gap. From a band theory perspective, we confirm that curvature serves as a critical tuning parameter, capable of driving topological phase transitions by inducing band inversion.

## 1 Introduction

Topological materials represent a new quantum state of matter in condensed matter physics known as the topological phase, which cannot be classified by Landau's symmetry-breaking theory. Instead, it is characterized by protected edge or surface states originating from the non-trivial topological character of the bulk wave functions. From a band theory perspective, a topological phase remains invariant as long as the Hamiltonian can be continuously deformed without closing the energy gap. The prototypical example of this is the integer quantum Hall effect [1], where hall conductivity is quantized in terms of the TKNN invariant [2]. This invariant is fundamentally linked to the Berry phase of Bloch states in momentum space, establishing a bulk-boundary correspondence between the number of occupied Landau levels and the count of edge states. In 2005, Bernevig et al. [3] proposed the topological insulator, where strong spin-orbit coupling (SOC) triggers a band inversion process, leading to gapless edge states protected by time-reversal symmetry. Subsequently, Kane and Mele [4] demonstrated that topological insulators are classified by the $Z_2$ invariant, associated with time-reversal and inversion symmetries. These developments gave birth to the concept of symmetry-protected topological (SPT) phases, prompting physicists to generalize the relationship between topological phases and discrete symmetries across all spatial dimensions, as evidenced by the periodic table classification of non-interacting fermion systems proposed by Kitaev [5] and Altland [6].

A prominent example of topological phases in strongly correlated systems is the topological superconductor [7][8]. The edges or surfaces of these superconductors can host zero-energy Majorana fermions [9]—quasiparticles that are their own antiparticles. These fermions obey non-Abelian statistics, making them ideal candidates for fault-tolerant topological quantum computing qubits [10,11]. In superconductors, topological phases typically require triplet pairing. Theoretical models suggest that p-wave triplet pairing [11,13] can be induced by strong SOC in hybrid structures [14,15] or in non-centrosymmetric superconductors (NCSs) [16-18]. In NCSs, the absence of inversion symmetry, often caused by Rashba SOC, leads to a mixture of singlet and triplet pairing states. Here, the topological phase transition is governed by the competition between singlet and triplet potentials. Crucially, such systems are inherently protected by particle-hole symmetry, which ensures that the excitation spectrum is symmetric around zero energy—a prerequisite for the existence of Majorana modes.

Beyond tuning intrinsic material parameters, geometric curvature has been identified as a powerful degree of freedom to manipulate quantum states. In a seminal work, da Costa [19] demonstrated that constraining a quantum particle to a curved surface induces an attractive geometric potential, $V_g$, driven solely by the surface mean curvature. This framework was later extended to systems with spin-orbit coupling (SOC), where the interplay between geometry and spin dynamics becomes prominent. Chang et al. [20] derived the effective Hamiltonian for Rashba and Dresselhaus SOC on curved surfaces, revealing that curvature renormalizes the SOC strength and induces curvature-dependent anisotropy in the spin texture. More importantly, this geometric reconfiguration of SOC offers robust protection for spin transport. Tan et al. [21] demonstrated the resilience of spin-orbit torque (SOT) against geometrical backscattering in curvilinear nanostructures. They showed that as electrons propagate along a curved manifold, the local SOC field rotates adiabatically with the surface normal, suppressing spin-flip scattering caused by geometric defects. Furthermore, recent studies on electron-hole systems by Siu et al. [22] confirmed that periodic curvature can induce band inversion and partner switching, establishing curvature as a decisive driver for topological phase transitions without altering the material's chemical composition.

Inspired by these findings, we postulate that curvature plays a similarly critical role in the Bogoliubov-de Gennes (BdG) spectrum of topological superconductors. In NCSs, the topological phase is dictated by the ratio of the SOC-induced triplet pairing to the singlet pairing. Since curvature fundamentally modifies the effective SOC and introduces a geometric potential that shifts the effective chemical potential, it serves as a natural tuning parameter to modulate this ratio. In this paper, we extend these geometric concepts to s+p wave NCS systems on a 2D cylindrical manifold. By solving the BdG eigenfunctions analytically, we determine the critical relationship between surface curvature (defined by the radius $\rho$) and the phase transition boundaries. The paper is organized as follows: In Section 2, we derive the effective BdG Hamiltonian for the cylindrical surface. In Section 3, we obtain analytical expressions for the energy spectrum. Section 4 provides a numerical analysis of curvature-driven band inversion, and Section 5 concludes our findings.

*Corresponding author
✉ harvey910344@gmail.com

## 2 MODEL HAMILTONIAN

### A. Hamiltonian in square lattice

In a noncentrosymmetric system, the absence of an inversion center allows for a momentum-dependent spin-orbit coupling (SOC). We consider a two-dimensional $s + p$ wave noncentrosymmetric superconductor (NCS) on a square lattice. The system's crystal symmetry is described by the $\mathrm{C_{4v}}$ point group, which dictates the form of the allowed SOC vector field $\mathrm{g_k}$. Under this symmetry, the leading-order SOC vector near the Γ point transforms as the irreducible representation of $\mathrm{C_{4v}}$, given by $\mathrm{g_k} = \alpha(k_y\hat{x} - k_x\hat{y})$, where $\alpha$ is the Rashba coupling constant. We describe the system using the Bogoliubov-de Gennes (BdG) formalism. The Nambu spinor basis is defined as $\Psi_\mathrm{k} = (c_{\mathrm{k}\uparrow}, c_{\mathrm{k}\downarrow}, c^+_{-k\uparrow}, c^+_{-k\downarrow})^T$. The effective single-particle Hamiltonian is given by:

The effective single-particle Hamiltonian is given by:

$$H_{BdG}(\boldsymbol{k}) = \begin{pmatrix} H_0(\boldsymbol{k}) & \Delta(\boldsymbol{k}) \\ \Delta^+(k) & -H_0^T(-\boldsymbol{k}) \end{pmatrix} \tag{1}$$

where $H_0(\boldsymbol{k})$ represents the normal state Hamiltonian including the kinetic energy and the Rashba SOC:

$$H_0(\boldsymbol{k}) = \xi_{\boldsymbol{k}}\sigma_0 + \alpha(k_y\sigma_x - k_x\sigma_y) \tag{2}$$

Here, $\xi_\mathbf{k} = \frac{\hbar^2\mathbf{k}^2}{2m} - \mu$ is the single-particle energy measured relative to the chemical potential $\mu$.

The pairing potential $\Delta(\boldsymbol{k})$ consists of a mixture of spin-singlet s-wave and spin-triplet p-wave components. Following the symmetry of the Rashba vector, the triplet d-vector is aligned parallel to $\mathrm{g_k}$, yielding

$$\boldsymbol{\Delta}(\mathbf{k}) = (\Delta_\mathrm{s}\sigma_0 + \Delta_\mathrm{p}\mathrm{g_k} \cdot \sigma/\alpha)\mathrm{i}\sigma_\mathrm{y} \tag{3}$$

By diagonalizing Eq. (1), the energy spectrum of the superconducting state in the flat square lattice is obtained as:

$$E_{\pm}^{2} = (\xi_{\mathbf{k}})^2 + (\alpha k)^2 + {\Delta_s}^2 + \Delta_p k^2 \pm 2\sqrt{k^2\left[\Delta_s\Delta_p + \alpha\xi_{\mathbf{k}}\right]^2} \qquad (4)$$

## B. Hamiltonian in cylindrical coordinate

When the 2D plane is rolled into a cylindrical manifold with radius $\rho$, the translational symmetry is changed, and the system must be described in generalized curvilinear coordinates. According to the thin-layer quantization procedure proposed by da Costa [19], the effective kinetic Hamiltonian on a curved surface is given by:

$$\widetilde{H_{kin}} = -\frac{\hbar^2}{2m^*}\frac{1}{\sqrt{g}}\partial_m\left(\sqrt{g}\,g^{mn}\partial_n\right) + V_g(q_1, q_2) \qquad (5)$$

The first term represents the Laplace-Beltrami operator, which is the kinetic energy in generalized coordinates, where indices m, n run over the surface coordinates, $g$ is the determinant of the metric tensor $g^{mn}$, and $m^*$ is the effective electron mass. The second term, $V_g$, is the geometric potential induced by the curvature effect. To explicitly define this potential, we introduce the Weingarten curvature tensor ($\alpha_j^i$). This tensor bridges the intrinsic geometry and the extrinsic bending of the surface by combining the metric tensor (the First Fundamental Form, which describes the distribution of surface coordinates) with the curvature tensor (the Second Fundamental Form, which describes the degree of surface bending). The trace and determinant of the Weingarten matrix $\alpha_j^i$ correspond to the Mean Curvature (M) and Gaussian Curvature (K) of the surface, respectively. The geometric potential is thus expressed as:

$$V_g = -\frac{\hbar^2}{8m^*}\left(\mathrm{Tr}(\alpha)^2 - 4\mathrm{Det}(\alpha)\right) = -\frac{\hbar^2}{8m^*}(4M^2 - K) \qquad (6)$$

For a cylindrical surface of radius $\rho$, the Gaussian curvature is zero (K=0) and the Mean curvature is constant ($M = -\frac{1}{2}\rho$). Consequently, the geometric potential simplifies to a constant attractive term $V_g = -\frac{\hbar^2}{8m^*\rho^2}$.

For the spin-orbit coupling, we adopt the Uniform Rolling Model. By directly mapping the planar spin-momentum locking relationship onto the cylindrical geometry, we assume the effective electric field remains locally perpendicular to the surface. This reconfigures the Rashba interaction into the cylindrical basis $\{\widehat{\phi}, \hat{z}\}$ as:

$$\widetilde{H_{SOC}} = \frac{\alpha}{\hbar}\left(\sigma_z p_\phi - \sigma_\phi p_z\right) \tag{7}$$

Here, $\sigma_\phi = \left(-\sin\phi\,\sigma_x + \cos\phi\,\sigma_y\right)$ is the tangential spin operator rotating with the surface normal, and $p_\phi = -i\hbar\frac{1}{\rho}\partial_\phi$.

The p-wave pairing potential transforms in a similar manner, with the d-vector following the rotation of the local SOC field:

$$\widetilde{\Delta_p} = \frac{\Delta_p}{\hbar}\left(\sigma_z p_\phi - \sigma_\phi p_z\right) \tag{8}$$

Finally, substituting the cylindrical forms of the kinetic energy (with $V_g$) and the SOC term into the BdG formalism, we obtain the effective Hamiltonian for the curved superconductor:

$$\widetilde{H_{cyl}}(\boldsymbol{k}) = \begin{pmatrix} \frac{\hbar^2}{2m^*}(\partial_z^2 + \frac{1}{\rho^2}\partial_\phi^2) + V_g - \mu + \widetilde{H_{SOC}} & \left(\Delta_s + \widetilde{\Delta_p}\right)i\sigma_y \\ \left[\left(\Delta_s + \widetilde{\Delta_p}\right)i\sigma_y\right]^+ & -\left[\frac{\hbar^2}{2m^*}(\partial_z^2 + \frac{1}{\rho^2}\partial_\phi^2) + V_g - \mu + \widetilde{H_{SOC}}\right]^* \end{pmatrix} \tag{9}$$

This Hamiltonian (Eq. 9) serves as the basis for our subsequent analysis of the curvature-induced topological phase transitions.

## 3 Solution Of The Bogoliubov-De Gennes Hamiltonian In Cylindrical Coordinate

Since the effective Hamiltonian on the cylindrical manifold (Eq. 9) preserves the translational symmetry along the z-axis and the rotational symmetry about the z-axis (in our Uniform Rolling Model), the longitudinal momentum $k_z$ and the total angular momentum $j_z$ are conserved quantities. We can thus propose the ansatz for the eigenspinors as:

$$\Psi = \frac{1}{\sqrt{2\pi L}}\begin{pmatrix} e^{i\left(n-\frac{1}{2}\right)\varphi} \\ e^{i\left(n+\frac{1}{2}\right)\varphi} \\ e^{i\left(n+\frac{1}{2}\right)\varphi} \\ e^{i\left(n-\frac{1}{2}\right)\varphi} \end{pmatrix} e^{i\frac{p_z}{\hbar}z} \tag{10}$$

Here, $k_z$ is the continuous wave vector along the cylinder axis, and n is the quantized angular momentum quantum number. Due to the anti-periodic boundary condition required for spinors upon a $2\pi$ rotation, n takes **half-integer values** ($n = \pm 1/2, \pm 3/2, \ldots$). The phase factors $e^{i\phi}$ in the spinor components arise from the rotation of the local spin basis $\{\hat{\phi}, \hat{z}, \hat{\rho}\}$ relative to the global frame, ensuring the single-valuedness of the wavefunction.

Substituting this ansatz into the BdG equation $\widetilde{H_{cyl}}\Psi = E\Psi$, the problem reduces to diagonalizing a $4 \times 4$ matrix Hamiltonian $H_{n,k_z}$ in the basis of wavefunction. The matrix elements are determined by the effective kinetic energy and SOC terms derived in the previous section:

$$H_{n,k_z} = \begin{pmatrix} \boldsymbol{h_z + h_1} & \mathbf{i\alpha k_z} & \mathbf{-i\Delta_p k_z} & \Delta_s + \frac{\Delta_p}{\rho}\left(\mathbf{n} - \frac{1}{2}\right) \\ \mathbf{-i\alpha k_z} & \boldsymbol{h_z + h_2} & -\Delta_s + \frac{\Delta_p}{\rho}\left(\mathbf{n} + \frac{1}{2}\right) & \mathbf{-i\Delta_p k_z} \\ \mathbf{i\Delta_p k_z} & -\Delta_s + \frac{\Delta_p}{\rho}\left(\mathbf{n} + \frac{1}{2}\right) & \boldsymbol{-h_z - h_2} & \mathbf{-i\alpha k_z} \\ \Delta_s + \frac{\Delta_p}{\rho}\left(\mathbf{n} - \frac{1}{2}\right) & \mathbf{i\Delta_p k_z} & \mathbf{i\alpha k_z} & \boldsymbol{-h_z - h_1} \end{pmatrix} \tag{11}$$

Where $\boldsymbol{h_z}$, $\boldsymbol{h_1}$, $\boldsymbol{h_2}$ denote $\frac{\hbar^2 \mathbf{k_z^2}}{2m^*} - \boldsymbol{\mu}$, $\left(\frac{\hbar^2}{2m^*\boldsymbol{\rho}^2}\right)(\mathbf{n^2 - n}) + \frac{\boldsymbol{\alpha}}{\boldsymbol{\rho}}\left(\mathbf{n} - \frac{1}{2}\right)$, $\left(\frac{\hbar^2}{2m^*\boldsymbol{\rho}^2}\right)(\mathbf{n^2 + n}) - \frac{\boldsymbol{\alpha}}{\boldsymbol{\rho}}\left(\mathbf{n} + \frac{1}{2}\right)$

respectively.

By solving the characteristic equation $\det\left(H_{n,k_z} - EI\right) = 0$, we obtain the analytical expression for the energy square of the quasiparticle excitations:

$$E^2 = \frac{1}{2}\left[\frac{\hbar^2}{2m^*\rho^2}(n^2+n) + \left(\frac{\hbar^2 {k_z}^2}{2m^*} - \mu\right) - \frac{\alpha}{\rho}\left(n+\frac{1}{2}\right)\right]^2 + +\frac{1}{2}\left[\frac{\hbar^2}{2m^*\rho^2}(n^2-n) + \left(\frac{\hbar^2 {k_z}^2}{2m^*} - \mu\right) + \frac{\alpha}{\rho}\left(n-\frac{1}{2}\right)\right]^2 + \frac{1}{2}\left\{\left[-\Delta_s + \frac{\Delta_p}{\rho}\left(n+\frac{1}{2}\right)\right]^2 + \left[\Delta_s + \frac{\Delta_p}{\rho}\left(n-\frac{1}{2}\right)\right]^2\right\} + \left(\Delta_p + \alpha^2\right)k_z^2 \pm \frac{1}{2}\sqrt{D} \tag{12}$$

Where The discriminant D encapsulates the complex interplay between the Rashba SOC, the singlet-triplet pairing competition, and the curvature-induced energy shifts, and is defined as:

$$D = \left\{\left[\frac{\hbar^2}{2m^*\rho^2}(n^2+n) + \left(\frac{\hbar^2 {k_z}^2}{2m^*} - \mu\right) - \frac{\alpha}{\rho}\left(n+\frac{1}{2}\right)\right]^2 - \left[\frac{\hbar^2}{2m^*\rho^2}(n^2-n) + \left(\frac{\hbar^2 {k_z}^2}{2m^*} - \mu\right) + \frac{\alpha}{\rho}\left(n-\frac{1}{2}\right)\right]^2 + \left[-\Delta_s + \frac{\Delta_p}{\rho}\left(n+\frac{1}{2}\right)\right]^2 - \left[\Delta_s + \frac{\Delta_p}{\rho}\left(n-\frac{1}{2}\right)\right]^2\right\}^2$$

$$+4k_z^2\left\{\alpha\left[\frac{\hbar^2}{2m^*\rho^2}(n^2+n)+\left(\frac{\hbar^2 {k_z}^2}{2m^*}-\mu\right)-\frac{\alpha}{\rho}\left(n+\frac{1}{2}\right)\right]+\alpha\left[\frac{\hbar^2}{2m^*\rho^2}(n^2+n)+\left(\frac{\hbar^2 {k_z}^2}{2m^*}-\mu\right)-\frac{\alpha}{\rho}\left(n+\frac{1}{2}\right)\right]+2\Delta_s\Delta_p\right\}^2$$

$$+4k_z^2\left\{\Delta_p\left[\frac{\hbar^2}{2m^*\rho^2}(n^2+n)+\left(\frac{\hbar^2 {k_z}^2}{2m^*}-\mu\right)-\frac{\alpha}{\rho}\left(n+\frac{1}{2}\right)\right]-\Delta_p\left[\frac{\hbar^2}{2m^*\rho^2}(n^2-n)+\left(\frac{\hbar^2 {k_z}^2}{2m^*}-\mu\right)+\frac{\alpha}{\rho}\left(n-\frac{1}{2}\right)\right]+\right.$$
$$\left.2\Delta_p\frac{n\alpha}{\rho}\right\}^2$$

$$-8\left\{\left[\frac{\hbar^2}{2m^*\rho^2}(n^2+n)+\left(\frac{\hbar^2 {k_z}^2}{2m^*}-\mu\right)-\frac{\alpha}{\rho}\left(n+\frac{1}{2}\right)\right]+\left[\frac{\hbar^2}{2m^*\rho^2}(n^2-n)+\left(\frac{\hbar^2 {k_z}^2}{2m^*}-\mu\right)+\frac{\alpha}{\rho}\left(n-\frac{1}{2}\right)\right]\right\}\alpha\left(\frac{\Delta_p}{\rho}\right)\Delta_p k_z^2$$

$$+4\left(\frac{\Delta_p}{\rho}\right)^2{\Delta_p}^2k_z^2-16\Delta_s\left(\frac{\Delta_p}{\rho}\right){\Delta_p}^2k_z^2 \tag{13}$$

In the following section, we will use this energy spectrum to numerically analyze the curvature-induced band inversion and the resulting topological phase transitions.

# 4 RESULT

In this section, we perform a numerical analysis of the quasiparticle excitation spectrum to determine the topological phase boundaries. According to Eq. (12), a topological phase transition is characterized by the closing and reopening of the superconducting energy gap at the high-symmetry points of the Brillouin zone. For our cylindrical system, we focus on the lowest energy excitations by fixing the angular momentum quantum number to $\mathrm{n}=\pm\frac{1}{2}$. We investigate the gap-closing condition (E = 0) by treating the p-wave pairing potential $\Delta_\mathrm{p}$ and the cylinder radius $\rho$ as the primary variables. To provide a concrete comparison, we set the physical parameters as follows: effective electron mass $m^*=\mathrm{m_e}$, chemical potential $\mu=0.125$ eV, and lattice constant $\mathrm{a}=1$ nm . We solve for the critical p-wave potential $\Delta_p$ that satisfies the gap-closing condition across a range of radii $\rho$, considering different values of the s-wave pairing potential $\Delta_\mathrm{s}$and Rashba SOC constant $\alpha$.

In $\mathrm{n}=\pm\frac{1}{2}$, The energy square in equation (12) is equal to

$$E^2=\frac{1}{2}\left[\frac{5\hbar^4}{32{\mathrm{m_e}}^2\rho^4}+2\left(\frac{\hbar^2k_z^2}{2\mathrm{m_e}}-\mu\right)+\left(\frac{\alpha}{\rho}\right)^2+\frac{1}{2}\left(\frac{\hbar^2{k_z}^2}{2\mathrm{m_e}}-\mu\right)-\frac{3\hbar^2}{4\mathrm{m_e}\rho^2}\left(\frac{\alpha}{\rho}\right)-2\left(\frac{\hbar^2k_z^2}{2\mathrm{m_e}}-\mu\right)\frac{\alpha}{\rho}\right]$$

$$+\frac{1}{2}\left[2{\Delta_\mathrm{s}}^2-2\Delta_\mathrm{s}\frac{\Delta_p}{\rho}+\left(\frac{\Delta_p}{\rho}\right)^2\right]+\left(\Delta_p^2+\alpha^2\right)k_z^2\pm\frac{1}{2}\sqrt{D_1} \tag{14}$$

$$D_1=\left[\left(\frac{\hbar^4}{8{\mathrm{m_e}}^2\rho^4}\right)+\left(\frac{\alpha}{\rho}\right)^2+\frac{\hbar^2}{\mathrm{m_e}\rho^2}\left(\frac{\hbar^2{k_z}^2}{2\mathrm{m_e}}-\mu\right)-\frac{3\hbar^2}{4\mathrm{m_e}\rho^2}\left(\frac{\alpha}{\rho}\right)-2\left(\frac{\hbar^2{k_z}^2}{2\mathrm{m_e}}-\mu\right)\left(\frac{\alpha}{\rho}\right)-2\Delta_\mathrm{s}\frac{\Delta_p}{\rho}+\left(\frac{\Delta_p}{\rho}\right)^2\right]^2$$
$$4k_z^2\left\{\alpha\left[\frac{\hbar^2}{4\mathrm{m_e}\rho^2}+2\left(\frac{\hbar^2{k_z}^2}{2\mathrm{m_e}}-\mu\right)-\frac{\alpha}{\rho}\right]+2\Delta_\mathrm{s}\Delta_p\right\}^2+4k_z^2\left(\frac{\hbar^2}{2\mathrm{m_e}\rho^2}\right)^2\Delta_p^2-16\Delta_\mathrm{s}\frac{\Delta_p}{\rho}\Delta_p^2{k_z}^2+4\left(\frac{\Delta_p}{\rho}\right)^2\Delta_p^2k_z^2$$
$$-8\left[\frac{\hbar^2}{4\mathrm{m_e}\rho^2}+2\left(\frac{\hbar^2{k_z}^2}{2\mathrm{m_e}}-\mu\right)-\frac{\alpha}{\rho}\right]\alpha\frac{\Delta_p}{\rho}\Delta_p k_z^2 \tag{15}$$

The numerical results are illustrated in Fig. 1 and Fig. 2. Fig. 1 presents the phase boundary for the case of $\Delta_s = 3$ meV with three different Rashba coupling strengths. As expected, in the limit of a large radius ($\rho \to \infty$), the critical value for the topological phase transition converges to the result of a flat square lattice. This is analytically consistent with Eq. (14), where the curvature-dependent terms $n/\rho$ and $V_g$ become negligible as $\rho$ increases. However, as the radius $\rho$ decreases, the curvature effect becomes significant. We observe that the Rashba constant $\alpha$ plays a dual role in modulating the phase transition. In small $\alpha$ case, the Rashba coupling is weak, the region where the topological phase transition is suppressed dominates. In this regime, the curvature acts as a barrier, requiring a larger $\Delta_p$ to achieve band inversion compared to the flat lattice. In Large $\alpha$ case, conversely, with a larger Rashba constant, the enhancement of the topological phase transition becomes prominent. The interplay between the reconfigured SOC and the geometric potential lowers the threshold for band inversion, making the topological phase more accessible at smaller radii.

Fig. 2 shows the results for a stronger s-wave pairing potential, $\Delta_s = 7$ meV. While the general trends regarding the influence of $\alpha$ and $\rho$ remain similar to Fig. 1, a fascinating observation is that the efficiency of the curvature-induced enhancement is more conspicuous at higher $\Delta_s$. This suggests that systems with stronger intrinsic pairing correlations are more sensitive to geometric tuning, allowing for a more effective control of Majorana modes through mechanical deformation of the nanostructure.

## 5 Summary And Conclusion

In this work, we have theoretically investigated the topological phase transitions of a two-dimensional s+p wave noncentrosymmetric superconductor (NCS) on a curved cylindrical manifold. By employing the thin-layer quantization formalism and the Uniform Rolling Model, we derived an effective Bogoliubov-de Gennes (BdG) Hamiltonian that explicitly incorporates the da Costa geometric potential $V_g$ and the reconfiguration of Rashba spin-orbit coupling (SOC). Our model, rooted in the $C_{4v}$point group symmetry, provides a robust framework to describe how surface curvature modifies the quasiparticle excitation spectrum.

Our analytical and numerical results reveal that the cylinder radius $\rho$ (inverse curvature) serves as a decisive tuning parameter for topological phase transitions. In the limit of a large radius, the system's topological boundaries naturally converge to those of a flat square lattice. However, as the radius decreases, the interplay between the geometric potential and the curvature-modified SOC leads to significant shifts in the gap-closing conditions. Specifically, we observed a dual role played by the Rashba coupling strength $\alpha$. First, in regimes with weak SOC, the curvature-induced effects tend to suppress the topological phase, requiring a higher triplet pairing potential $\Delta_p$ to achieve band inversion. Conversely, in regimes with strong SOC and s-wave pairing, the geometric effects significantly enhance the stability of the topological phase, lowering the threshold for the transition.

These findings suggest that the topological properties and the resulting Majorana zero modes in curved nanostructures are highly sensitive to their geometric configuration. The ability to modulate topological phases through curvature offers a promising new avenue for the mechanical control of Majorana fermions in superconducting nanocircuits. Such geometric tuning could potentially enhance the resilience of topological qubits against local perturbations, providing a new platform for fault-tolerant topological quantum computation.

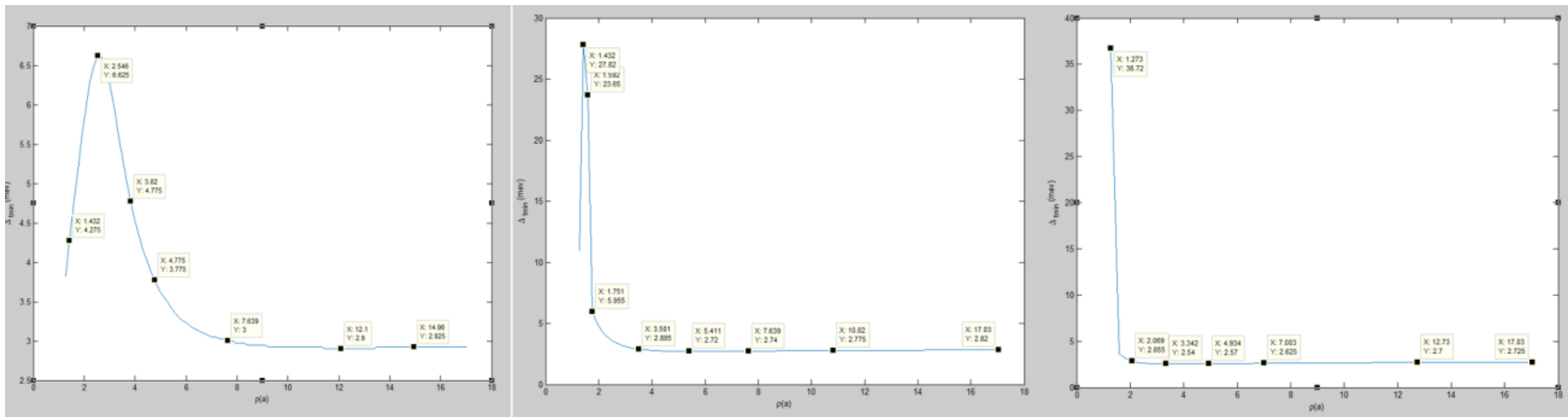

Figure 1: Left: the radius ρ(a) versus $\Delta_t$ in Δ=3mev, $\frac{\alpha}{a}$=0 mev case. Middle: the radius ρ(a) versus $\Delta_t$ in Δ=3mev, $\frac{\alpha}{a}$=8.5 mev case. Right: the radius ρ(a) versus $\Delta_t$in Δ=3mev, $\frac{\alpha}{a}$=17 mev case.

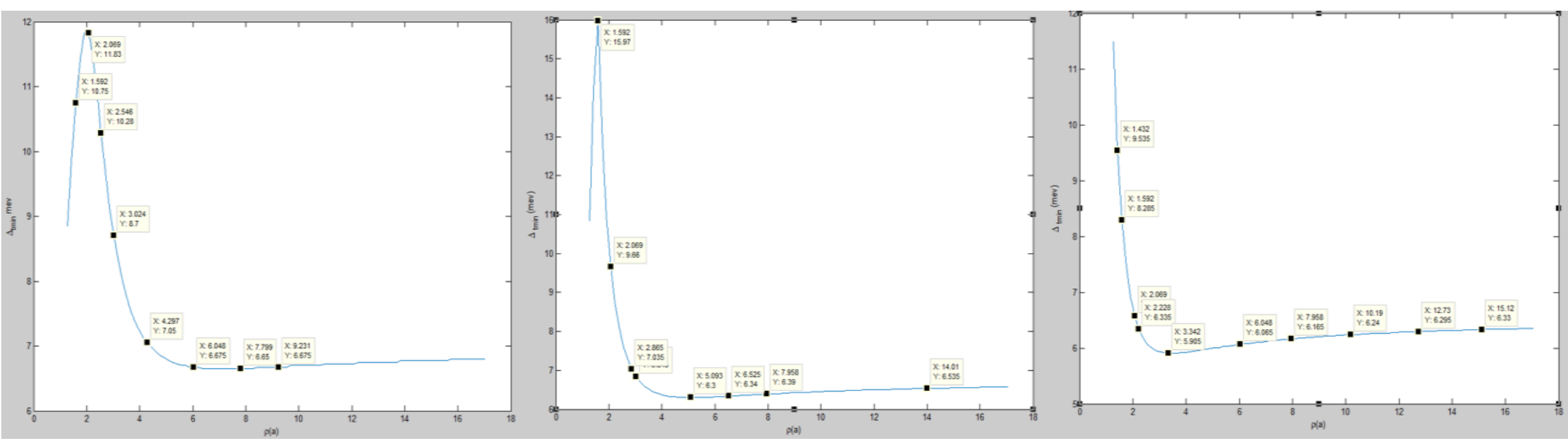

Figure 2: Left: the radius ρ(a) versus $\Delta_t$ in Δ=7mev, $\frac{\alpha}{a}$=0 mev case. Middle: the radius ρ(a) versus $\Delta_t$ in Δ=7mev, $\frac{\alpha}{a}$=8.5 mev case. Right: the radius ρ(a) versus $\Delta_t$ in Δ=7mev, $\frac{\alpha}{a}$=17 mev case.